\documentclass[12pt]{article}
\usepackage{latexsym}
\usepackage{graphicx, amssymb, epsfig}
\usepackage{amsmath}
\usepackage{empheq}
\usepackage{subfigure}

\newcommand{\trace}[1]{\mbox{Tr}\left({#1}\right)}
\newcommand{\pr}[1]{\mbox{Pr}\left\{{#1}\right\}}
\newcommand{\spa}[1]{\textrm{$\sf{sp}$}\left\{{#1}\right\}}
\newcommand{\ud}{\mathrm{d}}

\newtheorem{lemma}{Lemma}

\newtheorem{remark}{Remark}

\begin{document}


\title{Opportunistic Spatial Orthogonalization and Its Application in Fading Cognitive Radio Networks}

\author{\normalsize
Cong Shen
\and
\normalsize Michael P. Fitz \footnote{Part of the material in this paper was presented at the 43rd Conference on Information Sciences and Systems (CISS), Baltimore, MD, Mar. 2009. Cong Shen and Michael P. Fitz are with the Department of Electrical Engineering, University of California Los Angeles (UCLA), Los Angeles, CA 90095, USA. Email: \texttt{\{congshen,fitz\}@ee.ucla.edu}. Michael P. Fitz is also with Northrop Grumman Space Technology, Redondo Beach, CA 90278, USA.}}


\date{\today}
\maketitle

\begin{abstract}
Opportunistic Spatial Orthogonalization (OSO) is a  cognitive radio scheme that allows the existence of secondary users and hence increases the system throughput, even if the primary user occupies all the frequency bands all the time. Notably, this throughput advantage is obtained without sacrificing the performance of the primary user, if the interference margin is carefully chosen. The key idea is to exploit the spatial dimensions to orthogonalize users and hence minimize interference. However, unlike the time and frequency dimensions, there is no universal basis for the set of all multi-dimensional spatial channels, which motivated the development of OSO. On one hand, OSO can be viewed as a \emph{multi-user diversity} scheme that exploits the channel randomness and independence. On the other hand, OSO can be interpreted as an \emph{opportunistic interference alignment} scheme, where the interference from multiple secondary users is opportunistically aligned at the direction that is orthogonal to the primary user's signal space. In the case of multiple-input multiple-output (MIMO) channels, the OSO scheme can be interpreted as ``riding the peaks'' over the eigen-channels, and ill-conditioned MIMO channel, which is traditionally viewed as detrimental, is shown to be \emph{beneficial} with respect to the sum throughput. Throughput advantages are thoroughly studied, both analytically and numerically.
\end{abstract}

\section{Introduction}
\label{sec:intro}

The focus of this work is on the multi-user interference problem in a cognitive radio (CR) environment. The central problem in general wireless networks is the \emph{multi-user interference}. The main technique that mitigates this problem is to orthogonalize  users onto different degrees of freedom to minimize interference among them\footnote{Although lately there has been intensive research on power-splitting schemes (e.g., \cite{ETW:08}) first proposed by Han and Kobayashi \cite{HK:81}, it is still immature in practice and the overwhelming technique remains to be orthogonalization.}. Such orthogonalization is traditionally done in the time or frequency domain. The development of CR is one such example. The essential idea of CR is that some frequency bands may be unoccupied by the primary user (PU) for a long time, which causes a severe waste of the already scarce frequency resources. By allowing secondary users (SUs) to detect and transmit on these idle frequencies, resources are better utilized without sacrificing PU's performance, since users occupy the same frequency bands in different time and hence are orthogonal.

Ever since the development of point-to-point multiple-input multiple-output (MIMO) systems, it is well accepted that spatial dimensions can be exploited to provide degrees of freedom in addition to the usual time and frequency dimensions. It is also shown that the MIMO capacity, either in a point-to-point system or a multi-user system, can be significantly increased by the additional spatial dimensions. A natural question is whether the user orthogonalization principle can be extended to the spatial domain for non-centralized wireless networks\footnote{For centralized networks such as cellular, SDMA orthogonalizes users in the spatial domain.}. This has important advantages in a CR system. As we have mentioned, throughput improvement of CR relies on the fact that PU is not using all the frequency bands all the time, i.e., there are some ``spectral holes'' for SUs to exploit. This observation breaks down in some applications where PU is intensively active, and hence the frequency holes are very few and difficult to find. As a result, traditional CR will allow very few SUs to transmit in order to maintain the communication quality of the primary link (this is denoted as a ``PU only'' scheme). In this case, exploiting spatial dimensions would be able to accommodate SUs in such a way that the overall throughput is increased without causing intolerable interference to the PU.

However, there is one fundamental difference between the spatial and the time/frequency dimensions: there is no universal basis for the set of all multi-dimensional spatial channels. This is in sharp contrast to both the time domain, where any signal can be orthogonalized to different time units, and the frequency domain, where the overall signal can be decomposed (by the FFT operation in practice) into components on different subcarriers and made orthogonal to each other if the frequency separation is chosen properly. The set of all MIMO channel matrices, however, is lack of such universal basis and each matrix will have its own coordinate (singular vectors) and length (singular values). In other words, it is impossible to diagonalize all channel matrices onto a given set of orthonormal basis and assign different users to different dimensions. Hence, exploiting the spatial dimensions in distributed wireless networks faces some conceptual difficulties.

There are some works attacking this problem. For example, Interference Alignment (IA) has been proposed \cite{CJ:08} to study the capacity of interference networks. The basic idea is to align interference from multiple users to lie within only a few dimensions at the intended receiver, and hence reduces its impact to the useful signal. This idea was further developed in \cite{PDLC:08}, where the interference is aligned to unused directions resulted from water-filling power allocation. However, the major problem of \cite{CJ:08,PDLC:08} is that the strategy works only if the SU knows the channel matrix of the primary link. Without this knowledge, the interference source node is unable to identify the unused directions, and hence does not know how to beam the signal. This assumption is highly unrealistic in practice, especially for a CR system, as there is no incentive for the PU to inform its primary channel to the potential SUs\footnote{Several iterative algorithms are proposed in \cite{GCJ:08} to remove the requirement of global channel knowledge, but the solutions rely on the \emph{reciprocity} of wireless networks (e.g., in TDD mode) to iteratively achieve interference alignment with only local channel knowledge.}. It is also for this reason that one should be very careful to formulate the resource optimization problem (e.g., maximize sum throughput or SU's throughput subject to the quality of service (QoS) constraint for PU), as this typically requires \emph{global} channel state information (CSI). For example, \emph{user scheduling} has been studied for the CR network in \cite{HZL:07,UN:09}. However, the optimization problem formulated in \cite{HZL:07} for joint beamforming and user scheduling requires global CSI, and the opportunistic scheduling policies in \cite{UN:09} focus on user dynamics from queuing theory.

The main contribution of this paper is the proposed OSO scheme, which allows both the PU and the SU(s) to transmit at the same time and the same frequency band, but use different spatial dimensions and hence the interference is  minimized. The spatial orthogonalization among users is achieved without requiring SU to know the primary link channel, but instead relying on exploiting the channel randomness and independence to take advantage of the multi-user diversity\footnote{It should be mentioned that exploiting multi-user diversity in cognitive radio networks has been studied in a recent work \cite{ZL:08}, but only single-antenna terminals are considered and hence it does not exploit spatial dimensions. In a different paper \cite{ZL2:08} multiple antennas were considered but not multi-user diversity. Also the interference from PU to SU was ignored.}. The OSO scheme is developed in great detail for the single-input multiple-output (SIMO) system, where the concept of opportunistic interference alignment is derived. The OSO scheme becomes more interesting in the MIMO setting where we show that ill-conditioned MIMO channel matrix, which is traditionally viewed as detrimental, in fact is beneficial as it allows ``riding the peaks'' to exploit multi-user diversity in CR.

Throughout the paper the following notations will be used. Matrices and vectors are denoted with bold capital and lowercase letters, respectively. $\mathbf{A}(i,j)$ is the $(i,j)$-th element of the matrix $\mathbf{A}$, and $\mathbf{a}(i)$ is the $i$-th element of the vector $\mathbf{a}$. $E_{X}[\cdot]$ denotes the expectation of a random variable with respect to the distribution of $X$. $a^*$ denotes the conjugate of the complex number $a$. $\langle\mathbf{a},\mathbf{b}\rangle \doteq \sum_i a(i)b(i)^*$ denotes the inner product of two complex vectors, and  $||\mathbf{a}|| \doteq \langle\mathbf{a},\mathbf{a}\rangle $ denotes the vector 2-norm of $\mathbf{a}$. For matrix, $\trace{\mathbf{A}}$ and $||\mathbf{A}||_F$ denote the trace and Frobenius norm of matrix $\mathbf{A}$, respectively. $\mathbf{A}^H$ is the Hermitian of the complex matrix $\mathbf{A}$.  Finally, we use $\spa{ \mathbf{v}_1, \cdots, \mathbf{v}_n}$ to denote the linear space spanned from vectors $\left\{ \mathbf{v}_1, \cdots, \mathbf{v}_n \right\}$.





%

\section{OSO for Cognitive Radio}
\label{sec:CR}

\subsection{System Model}
\label{sec:CR_model}

\begin{figure}
\centering
\includegraphics[width=\textwidth]{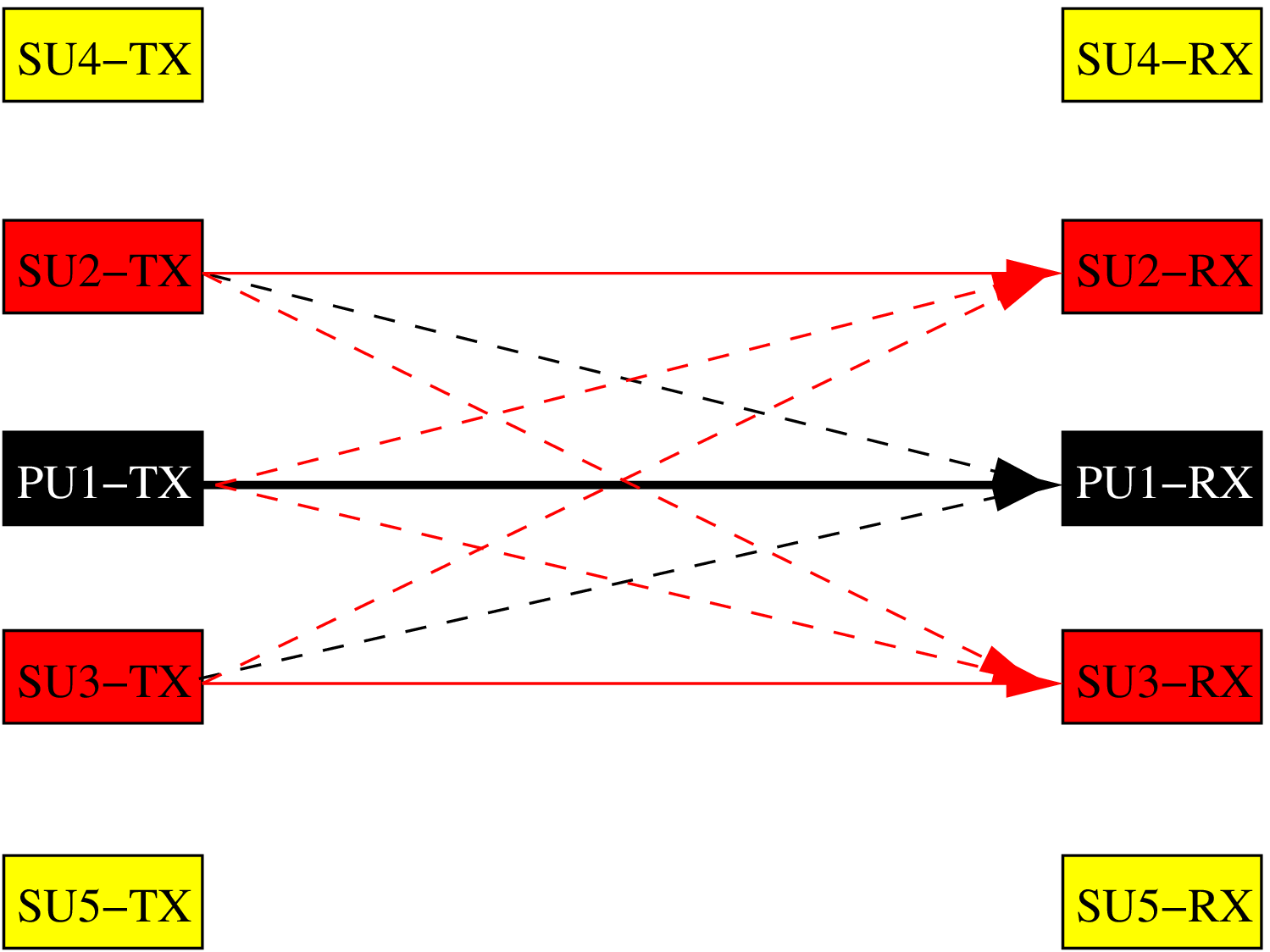}
\caption{An exemplary multi-user cognitive radio model with one primary link (PU1-TX to PU1-RX) and $K=4$ candidate secondary links. Among the candidate secondary links, $N=2$ are active (SU$n$-TX to SU$n$-RX, $n=2,3$), and the other two (SU$n$-TX to SU$n$-RX, $n=4,5$) are silent. Dashed black curves indicate the interference SUs cause to the primary link, which should be carefully controlled to guarantee PU's performance.}
\label{fig:CR_model}
\end{figure}

The system under consideration is illustrated in Fig.~\ref{fig:CR_model}. It is assumed that there is one primary link which occupies all the frequency bands and transmits all the time. There are $K$ candidate secondary links in the system, among which at most $N$ links are actively transmitting\footnote{The value of $N$ can be zero, which means that no SU is allowed to transmit. How to select these $N$ active links will be discussed in the sequel.}. We will index all the users together: User $1$(PU), $2, \cdots, N+1$ (SUs).


Consider that each link is equipped with $L_t$ transmit antennas and $L_r$ receive antennas. We use $\mathbf{H}_{i,j}$ to denote the channel between user $j$'s TX and user $i$'s RX, $i,j = 1, \cdots, N$. In the following discussion a special case of $L_t=1$ (SIMO) will be first discussed and then a general $(L_t, L_r)$ case is studied. In the former case $\mathbf{H}_{i,j}$ is in vector form and denoted as $\mathbf{h}_{i,j}$. Since both PU and SUs are allowed to transmit over all the frequency bands, our attention can be focused on a narrowband matrix interference channel:
\begin{equation}
\label{eqn:MIMO_N}
\begin{array}{rl}
\textrm{Primary receiver:} & \mathbf{y}_1 = \mathbf{H}_{1,1} \mathbf{x}_1 + \sum_{n=2}^{N+1} \mathbf{H}_{1,n} \mathbf{x}_n + \mathbf{z}_1; \\
\textrm{Secondary receivers:} & \mathbf{y}_m = \mathbf{H}_{m,m} \mathbf{x}_m + \sum_{n=1,n\neq m}^{N+1} \mathbf{H}_{m,n} \mathbf{x}_n + \mathbf{z}_m, \forall m=2, \cdots, N+1,
\end{array}
\end{equation}
where $\mathbf{x}_m$ and $\mathbf{y}_m$ are the transmit and receive vectors of the $m$-th user, respectively. We assume that PU$1$-TX has a total transmit power constraint $\trace{E[\mathbf{x}_1 \mathbf{x}_1^{H}]} \leq P_1$, and SU$m$-TX also has a total power constraint $\trace{E[\mathbf{x}_m \mathbf{x}_m^{H}]} \leq P_2$, $\forall m=2, \cdots, N+1$. Received signal at the $m$-th receiver is corrupted by an i.i.d. Additive White Gaussian Noise (AWGN) vector $\mathbf{z}_m \sim CN(\mathbf{0}, N_0 \mathbf{I})$.

The channel transfer matrices $\left\{ \mathbf{H}_{i,j} \right\}_{i,j=1}^{N+1}$ are assumed to be independent of each other. This assumption is generally valid if the transmitters/receivers are not co-located, which is the case for most CR systems. In all the numerical examples, it is further assumed that they are also identically distributed with each element $\mathbf{H}_{i,j}\left(l,m\right) \sim CN(0,1)$. This brings us to the familiar ground of i.i.d. MIMO Rayleigh fading channels. A block-fading model is assumed, i.e., the channel matrices are constant during the channel coherence time and changes independently to a different value according to some distribution. This channel model will be used to discuss the OSO scheme. We consider a coherent system where all the destination nodes have perfect knowledge of both the direct link from the intended source node and the cross links that come from the other source nodes, i.e., user $i$'s receiver knows $\left\{ \mathbf{H}_{i,j} \right\}_{j=1}^{N+1}$. Notice that $N$ is typically very small, which reduces the workload of channel estimation at SU-RXs. For the primary receiver (PU1-RX), we make a stronger assumption that it not only knows the direct link $\mathbf{H}_{1,1}$ but also has perfect knowledge of the channels that \emph{might} interfere with the primary link, i.e., $\left\{ \mathbf{H}_{1,j} \right\}_{j=2}^{K+1}$. This is necessary and can be achieved in the user selection stage, where all the $K$ candidate secondary users send orthogonal pilots and PU1-RX performs channel estimation for each cross-interference channel. In  applications such as uplink cellular networks being the PU (basestation is the PU-RX), this channel estimation can be reliably accomplished. Since all the destination nodes have perfect CSI, they can calculate the transmission rate that current channel state can support (taking into account the interference)  and feed it back to the source nodes\footnote{Note that this is not only a typical assumption in literature discussing multi-user diversity but also implemented in practice.}. This assumption eliminates the possibility of outage and allows us to use the average capacity as the performance measure.



A final note is that some assumptions are made only to simplify the sequel discussion. For example, we assume that the transmit/receive antennas are the same for each link, and there are only two power levels $P_1$ (PU) and $P_2$ (SUs). We also assume that the channel matrices are identically Gaussian distributed. These assumptions are not crucial to our discussion and can be relaxed to a more general situation. As for the assumptions of CSI, it is critical for PU1-RX to know $\left\{ \mathbf{H}_{1,j} \right\}_{j=2}^{K+1}$, as this is necessary for the selection of active SUs. However, the assumption that active SUs's destination nodes also know their cross-interference channels and feed back the instantaneous rate is made only to facilitate the use of average capacity. Without these assumptions, the proposed OSO scheme is still valid and one can consider the capacity versus outage as the performance measure.

\subsection{The OSO Scheme -- SIMO}
\label{sec:CR_OSO_SIMO}

One unique property of CR is the unbalance between users. Since PU has the legitimate right to operate on the given frequency bands, the SUs have to opportunistically sense the ``spectral holes'' and then transmit their own data without degrading the service of PU \cite{Haykin:05}. Due to this \emph{unbalanced} nature of CR, the system design is considerably different from the conventional multi-user network. The main difference is that the primary link is much more important than other secondary links, and hence the QoS of the primary link should be guaranteed, which means the interference caused by secondary transmissions needs to be carefully controlled. On the other hand, the QoS of secondary links is generally not guaranteed, which is a price SU has to pay when operating on a frequency band that is not licensed to him. Another consideration is that the primary link should be asked to make as few changes as possible to accommodate the existence of SUs. These design principles should be kept in mind when developing the OSO scheme for CR.

As has been discussed before, the major difficulty to exploit the spatial dimensions for cognitive radio is the lack of a universal basis for the spatial dimensions. Hence, it is impossible to diagonalize all possible channel matrices onto a given set of orthonormal basis and assign different users to different dimensions. The key idea of the proposed  solution is to exploit the \emph{channel randomness} and \emph{multi-user diversity} effect \cite{KH:95,VTL:02} to create multiple spatial dimensions and allow the primary and secondary users to occupy different dimensions to achieve (near) orthogonality. This idea is best illustrated with the SIMO system, which is the purpose of this section. For simplicity, we begin the discussion with $N=1$ (i.e., there is no more than one active SU) and generate to any $N$.

\subsubsection{$N=1$}
\label{sec:CR_OSO_SIMO_N1}

If one SU is present, the signal model (\ref{eqn:MIMO_N}) becomes
\begin{eqnarray}
\textrm{Primary link}&:& \mathbf{y}_1 = \mathbf{h}_{1,1} x_1 + \mathbf{h}_{1,2} x_2 + \mathbf{z}_1; \label{eqn:CR_OSO_SIMO_N1_PU} \\
\textrm{Secondary link}&:& \mathbf{y}_2 = \mathbf{h}_{2,2} x_2 + \mathbf{h}_{2,1} x_1 + \mathbf{z}_2. \label{eqn:CR_OSO_SIMO_N1_SU}
\end{eqnarray}
Since SU$1$-RX perfectly knows the channel gains $\mathbf{h}_{1,1}$ and $\mathbf{h}_{1,2}$, signal model (\ref{eqn:CR_OSO_SIMO_N1_PU}) is essentially a multiple-access channel (MAC). Complicated MAC decoders such as linear MMSE or interference cancellation can be employed. However, as one of our design principles is to make as few changes as possible, we would like to reuse the decoder when the interference does not exist. Such approach will make the receiver robust to the dynamic changes in the secondary transmissions. We would like to point out that in this work the principle is to treat interference as noise. In the interference channel literature, there are advanced techniques such as the combination of decoding the interference signal and treating interference as noise \cite{HK:81,ETW:08}. These techniques provide better throughput but are complicated to implement with today's technology.

Without interference model (\ref{eqn:CR_OSO_SIMO_N1_PU}) is a standard SIMO channel and maximum ratio combining (MRC) is the optimal decoder (i.e., matched filter):
\begin{eqnarray}
\label{eqn:CR_OSO_SIMO_N1_MRC}
\tilde{y}_1 &=& \langle \mathbf{y}_1,\mathbf{h}_{1,1}\rangle \nonumber \\
&=& || \mathbf{h}_{1,1} ||^2 x_1 + \langle \mathbf{h}_{1,2}, \mathbf{h}_{1,1} \rangle x_2 + \langle \mathbf{z}_1, \mathbf{h}_{1,1} \rangle.
\end{eqnarray}
Defining the \emph{interference power}
\begin{equation}
\label{eqn:CR_OSO_SIMO_N1_Beta_1}
\beta_{1}^{(1)} \doteq \frac{ |\langle \mathbf{h}_{1,2}, \mathbf{h}_{1,1} \rangle|^2}{|| \mathbf{h}_{1,1} ||^2}
\end{equation}
where the subscript denotes the number of candidate SUs and the superscript denotes the number of possible active SUs, the received signal-to-interference-plus-noise ratio (SINR) can be computed as
\begin{eqnarray}
\textrm{$\sf{SINR}_1$}(\beta_{1}^{(1)})  &=& \frac{ || \mathbf{h}_{1,1} ||^2 P_1 }{ \frac{ |\langle \mathbf{h}_{1,2}, \mathbf{h}_{1,1} \rangle|^2}{|| \mathbf{h}_{1,1} ||^2} P_2 +N_0}   \label{eqn:CR_OSO_SIMO_N1_MRC_SINR1} \\
&=& \frac{ || \mathbf{h}_{1,1} ||^2 P_1 }{ N_0 \left(  1 + \beta_{1}^{(1)} \frac{P_2}{N_0} \right) } \label{eqn:CR_OSO_SIMO_N1_MRC_SINR2} \\
 &\leq& \frac{ || \mathbf{h}_{1,1} ||^2 P_1 }{ N_0 } \label{eqn:CR_OSO_SIMO_N1_MRC_SNR3} \\
 &\doteq& \textrm{$\sf{SNR}_1$} \label{eqn:CR_OSO_SIMO_N1_MRC_SNR}
\end{eqnarray}
where $\textrm{$\sf{SNR}_1$}$ is the received signal-to-noise ratio of the primary link when there is no interference. The rate of the primary user is
\begin{equation}
C_1 = \log\left( 1+ \textrm{$\sf{SINR}_1$}(\beta_{1}^{(1)}) \right).
\end{equation}

\begin{figure}
\centering
\includegraphics[width=\textwidth]{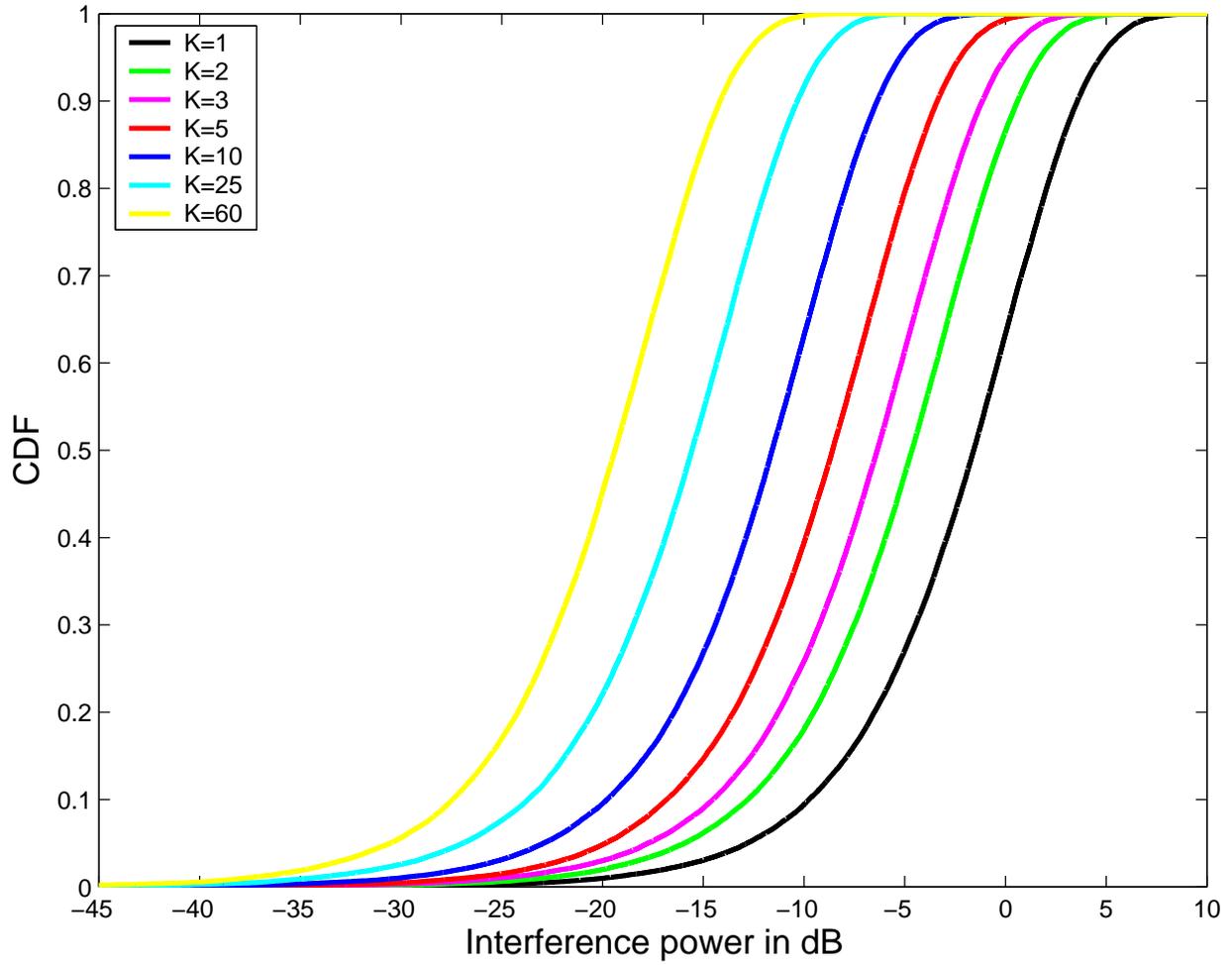}
\caption{CDF of the interference power $\beta_{K}^{(1)}$ for i.i.d. Gaussian SIMO channels with different values of $K$, $N=1$.}
\label{fig:plot2_CDF_dB}
\end{figure}

In order not to severely degrade the performance of PU, the interference power $\beta_{1}^{(1)}$ needs to be very small. In general this will not be true if $\mathbf{h}_{1,1}$ and $\mathbf{h}_{1,2}$ are independent. In Fig.~\ref{fig:plot2_CDF_dB} the curve corresponding to $K=1$ shows the Cumulative Distribution Function (CDF) of $\beta_{1}^{(1)}$ for the i.i.d. Gaussian channel.


However, when there are many candidate SUs in the system, there is a very good chance that PU1-RX can find one secondary link whose interference channel $\mathbf{h}_{1,k}$ is almost orthogonal to $\mathbf{h}_{1,1}$, and hence creates very little interference to the primary link. Notice that now the interference power becomes
\begin{equation}
\label{eqn:CR_OSO_SIMO_N1_Beta_K}
\beta_{K}^{(1)} \doteq \frac{ \min_{k=2, \cdots, K+1}|\langle \mathbf{h}_{1,k}, \mathbf{h}_{1,1} \rangle|^2}{|| \mathbf{h}_{1,1} ||^2}.
\end{equation}
The benefit of exploiting multiple candidate SUs can be analytically revealed by comparing (\ref{eqn:CR_OSO_SIMO_N1_Beta_1}) to (\ref{eqn:CR_OSO_SIMO_N1_Beta_K}). The distribution of $\beta_{K}^{(1)}$ with different values of $K$ is reported in Fig.~\ref{fig:plot2_CDF_dB}. As the number of candidate SUs increases in the system, the tail of the distribution for $\beta_{K}^{(1)}$ becomes much lighter. This in fact is due to the multi-user diversity effect. We name this scheme Opportunistic Spatial Orthogonalization (OSO).

As we can see from (\ref{eqn:CR_OSO_SIMO_N1_MRC_SNR3}), interference from SU reduces the SINR and the data rate of PU. We would like to perform the secondary user selection such that the interference power is below some threshold. Wireless fading channels have a wide range of dynamics and a constant QoS is very difficult to maintain. Typically the system design would assume a certain SINR margin to deal with the dynamics of channel fading and multi-user interference. We borrow this idea and define
\begin{eqnarray}
\gamma_{(dB)} &\doteq& \textrm{$\sf{SNR}_1$}_{(dB)} - \textrm{$\sf{SINR}_1$}(\beta_{K}^{(1)})_{(dB)} \label{eqn:CR_OSO_SIMO_N1_IntMar1} \\
&=& 10 \log_{10}\left( 1+ \beta_{K}^{(1)} \frac{P_2}{N_0}\right) \label{eqn:CR_OSO_SIMO_N1_IntMar2} \\
&\leq& \gamma_{\textrm{$\sf{thr}$}(dB)} \label{eqn:CR_OSO_SIMO_N1_IntMar3}
\end{eqnarray}
where $\gamma_{\textrm{$\sf{thr}$}(dB)}$ is the maximum interference threshold PU can tolerate. Hence, if a SU causes an interference that is no larger than this margin, it is allowed to use the same frequency band simultaneously with the PU. It should be noted that this interference threshold idea has already been discussed and adopted in the setting of CR spectrum sensing by the FCC as the \emph{interference temperature model} \cite{FCC:03}.

Now let us study the performance of SU. Similarly we insist that SU2-RX remains unchanged with and without interference. The MRC decoder at SU2-RX gives:
\begin{eqnarray}
\label{eqn:CR_OSO_SIMO_N1_MRC_SU}
\tilde{y}_2 &=& \langle \mathbf{y}_2,\mathbf{h}_{2,2}\rangle \nonumber \\
&=& || \mathbf{h}_{2,2} ||^2 x_2 + \langle \mathbf{h}_{2,2}, \mathbf{h}_{2,1} \rangle x_1 + \langle \mathbf{z}_2, \mathbf{h}_{2,1} \rangle,
\end{eqnarray}
and the corresponding SINR is
\begin{equation}
\label{eqn:CR_OSO_SIMO_N1_MRC_SINR_SU}
\textrm{$\sf{SINR}_2$} = \frac{ || \mathbf{h}_{2,2} ||^2 P_2 }{ N_0 + \frac{|\langle \mathbf{h}_{2,2}, \mathbf{h}_{2,1} \rangle|^2}{|| \mathbf{h}_{2,2} ||^2} P_1 }.
\end{equation}
Notice that the statistics to choose active SU is $\beta_{K}^{(1)}$, which is a random variable independent of $\textrm{$\sf{SINR}_2$}$. Hence, the performance of SU cannot always be guaranteed, due to the possibly strong interference the primary transmission could have caused. As is mentioned before, this is a price SU has to pay in order to be activated.

One might think of a joint user selection scheme, which first minimizes SU's interference to PU, and then selects the SU that also suffers minimum interference from the primary transmission. There are some practical problems with this scheme. First of all, such scheme requires two i.i.d. random variables $\frac{ |\langle \mathbf{h}_{1,1}, \mathbf{h}_{1,2} \rangle|^2}{|| \mathbf{h}_{1,1} ||^2}$ and $\frac{|\langle \mathbf{h}_{2,2}, \mathbf{h}_{2,1} \rangle|^2}{|| \mathbf{h}_{2,2} ||^2}$ to be small simultaneously. Although it is possible to show that asymptotically (as $K \rightarrow \infty$) this event will happen with probability one, the convergence rate is much slower because
\begin{eqnarray}
&{}&\pr{\frac{ |\langle \mathbf{h}_{1,1}, \mathbf{h}_{1,2} \rangle|^2}{|| \mathbf{h}_{1,1} ||^2} \leq \varepsilon, \frac{|\langle \mathbf{h}_{2,2}, \mathbf{h}_{2,1} \rangle|^2}{|| \mathbf{h}_{2,2} ||^2} \leq \varepsilon} \nonumber \\
&=& \pr{\frac{ |\langle \mathbf{h}_{1,1}, \mathbf{h}_{1,2} \rangle|^2}{|| \mathbf{h}_{1,1} ||^2} \leq \varepsilon}^2 \nonumber \\
&\ll& \pr{\frac{ |\langle \mathbf{h}_{1,1}, \mathbf{h}_{1,2} \rangle|^2}{|| \mathbf{h}_{1,1} ||^2} \leq \varepsilon}, \textrm{when $\varepsilon$ is very small.}
\end{eqnarray}
Another issue is that such scheme requires PU's receiver to have knowledge of all the interference channels $\left\{ \mathbf{h}_{k,1} \right\}_{k=2}^{K+1}$. This is a highly unrealistic assumption.



\begin{figure}
\centering
\includegraphics[width=\textwidth]{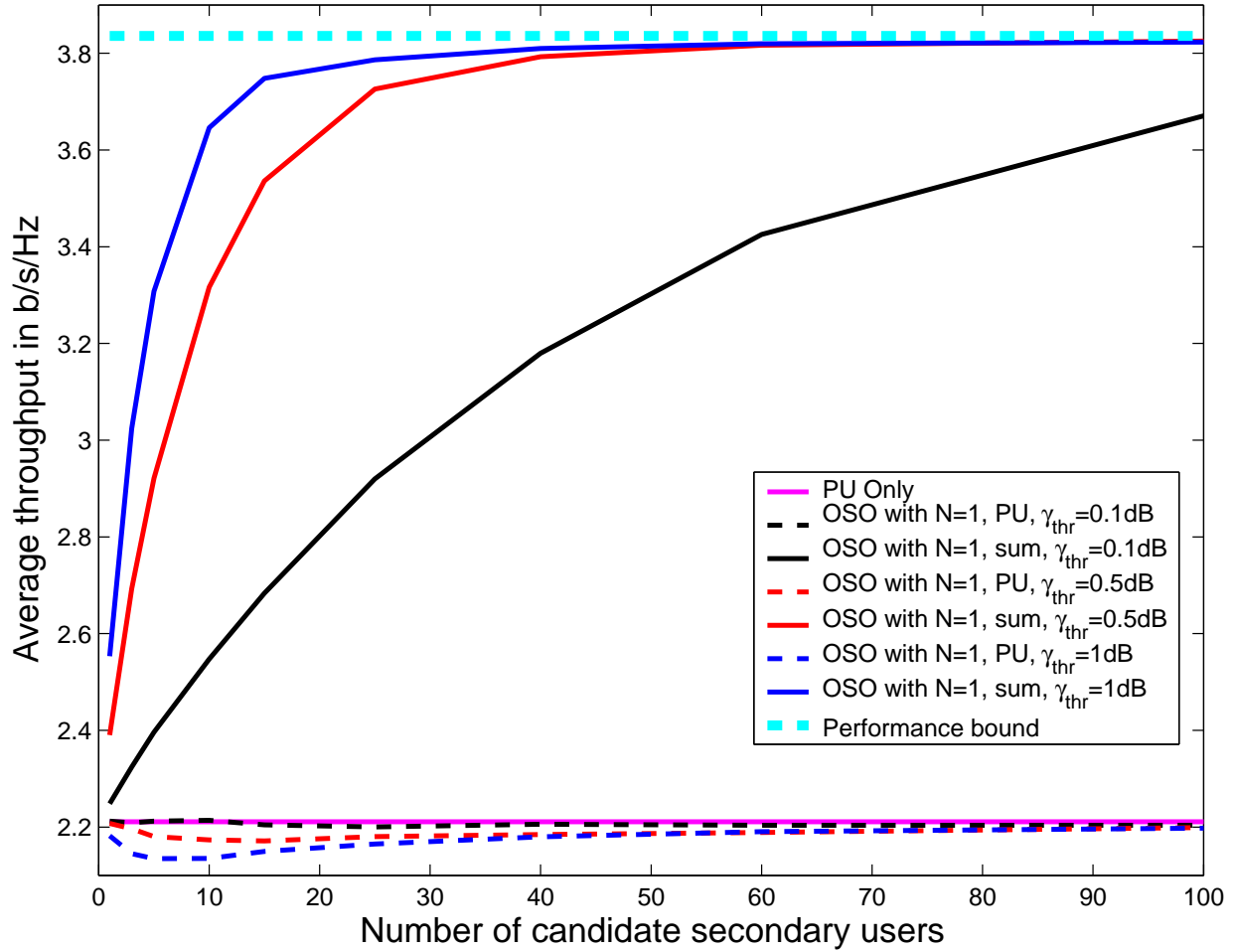}
\caption{Comparison of OSO and the conventional PU only scheme in a SIMO system with $L_r = 4$. Throughput versus number of candidate SUs is plotted. Maximum number of active SUs is $N=1$. Channels are i.i.d. Rayleigh fading. $\frac{P_1}{N_0} = \frac{P_2}{N_0} = 0$ dB. Three different interference threshold values are considered: $\gamma_{\textrm{$\sf{thr}$}} = 0.1, 0.5,$ and $1$ dB.}
\label{fig:plot1}
\end{figure}

Fig.~\ref{fig:plot1} gives a numerical example illustrating the benefits of OSO. A few interesting observations can be made from the figure. First of all, the performance degradation of PU due to the additional user is very little when the interference margin is stringent, but the overall throughput is increased significantly. When $\gamma_{\textrm{$\sf{thr}$}} = 0.1$ dB, the sum rate increase comes at almost no cost to the PU, whose throughput is nearly unchanged. Secondly, as the number of candidate SUs increases, the sum rate also increases. This is due to the multi-user diversity, where the probability that one candidate user is almost orthogonal to the primary link increases monotonically with the number of candidate users $K$. The third observation is that as $K$ becomes very large, the sum rate converges to a performance upper bound. This asymptotic behavior will be discussed in the sequel. Another observation is that when we relax the interference margin $\gamma_{\textrm{$\sf{thr}$}}$, PU may suffer a little more rate loss when $K$ is not very large (still below the margin), but the sum rate increases and converges to the upper bound much faster. This indicates that if the interference margin is not very stringent, only a small number of candidate SUs would allow the system to achieve optimal sum rate. For example, if $\gamma_{\textrm{$\sf{thr}$}} = 0.1$ dB then even with $K=100$ the sum rate is not saturated, but with $\gamma_{\textrm{$\sf{thr}$}} = 1$ dB only $K=40$ already approaches the upper bound. A final observation is as $K$ increases, PU's throughput in the OSO scheme first decreases and then increases. The initial throughput for PU is large because $K$ is very small, and with large probability none of the candidate SUs will be activated. As there are more and more candidate SUs, the probability that one of them will meet the interference margin requirement and hence be activated increases, which creates interference and reduces PU's throughput. As $K$ becomes very large, almost surely one SU is activated and thus brings  interference. However, due to the multi-user diversity effect, this interference will be extremely small when $K$ is very large, and PU's throughput will converge to the no-interference case.

Intuitively the asymptotic performance upper bound corresponds to when the primary link suffers no interference (SU is \emph{exactly} orthogonal to PU) and the throughput of both PU and SU is the ergodic capacity. The interference power $\beta_{K}^{(1)}$ is a non-increasing function of the number of candidate SUs $K$. In fact, we can prove the following lemma.
\begin{lemma}
\label{thm:betaK}
Consider independent and continuous random vectors $\left\{ \mathbf{h}_{1, k} \right\}_{k=1}^{K+1}$ with probability distribution functions (PDF) $f_k\left(  \mathbf{h} \right), k=1, \cdots, K+1$. Assume that the support  of $\mathbf{h}_{1, k}$ includes $\mathbf{0}^{+}$, i.e., $\int_{\mathbb{S}} f_k\left(  \mathbf{h} \right) \ud  \mathbf{h} > 0$ for any $\mathbb{S}$ that contains $\mathbf{0}^{+}$. Then $\forall \varepsilon > 0$ we have
\begin{equation}
\label{eqn:CR_OSO_SIMO_N1_thm_BetaK}
\pr{ \lim_{K\rightarrow \infty} \beta_{K}^{(1)} \leq \varepsilon } = 1
\end{equation}
with $\beta_{K}^{(1)}$ defined in (\ref{eqn:CR_OSO_SIMO_N1_Beta_K}).
\end{lemma}

\emph{Proof:}
It is equivalent to show that
\begin{equation}
\label{eqn:CR_OSO_SIMO_N1_thm_BetaK_proof1}
\pr{ \lim_{K\rightarrow \infty} \beta_{K}^{(1)} > \varepsilon } = 0.
\end{equation}
Define the event
\begin{equation}
A_k \doteq \left\{ \frac{|\langle \mathbf{h}_{1,k}, \mathbf{h}_{1,1} \rangle|^2}{|| \mathbf{h}_{1,1} ||^2} > \varepsilon \right\}.
\end{equation}
for $k=2, \cdots, K+1$. Then,
\begin{eqnarray}
\pr{ \lim_{K\rightarrow \infty} \beta_{K}^{(1)} > \varepsilon } &=& \pr{ \bigcap_{k=1}^{\infty} A_k } \\
&=& \int \pr{ \bigcap_{k=1}^{\infty} A_k | \mathbf{h}_{1, 1} }  f_1\left(  \mathbf{h}_{1, 1} \right) \ud \mathbf{h}_{1, 1} \\
&(a)\atop =& \int \prod_{k=1}^{\infty} \pr{ A_k | \mathbf{h}_{1, 1} }  f_1\left(  \mathbf{h}_{1, 1} \right) \ud \mathbf{h}_{1, 1} \label{eqn:CR_OSO_SIMO_N1_thm_BetaK_proof2}
\end{eqnarray}
where $(a)$ is due to the independence of $\left\{ A_k | \mathbf{h}_{1, 1} \right\}$. Now, since $\int_{\mathbb{S}} f_k\left(  \mathbf{h} \right) \ud  \mathbf{h} > 0$ for any $\mathbb{S}$ that contains $\mathbf{0}^{+}$, we have $ \pr{ A_k | \mathbf{h}_{1, 1} } < 1$ for any $\mathbf{h}_{1, 1} \neq \mathbf{0}$. Notice that $\pr{ \mathbf{h}_{1, 1} = \mathbf{0}} = 0$. Hence
\begin{equation}
\prod_{k=1}^{\infty} \pr{ A_k | \mathbf{h}_{1, 1} } \leq \lim_{K\rightarrow \infty} \left[ \max_{k}\pr{ A_k | \mathbf{h}_{1, 1} } \right]^{K} = 0
\end{equation}
which proves (\ref{eqn:CR_OSO_SIMO_N1_thm_BetaK_proof1}). This completes the proof.
\begin{flushright}
$\blacksquare$
\end{flushright}

With Lemma~\ref{thm:betaK}, we can proceed to study the asymptotic behavior of the OSO scheme. Since Lemma~\ref{thm:betaK} states that asymptotically the interference power $\beta_{K}^{(1)}$ goes to zero, the average throughput of PU will converge to the no-interference case, which is the ergodic capacity:
\begin{equation}
\label{eqn:CR_OSO_SIMO_N1_asympPU}
\lim_{K\rightarrow \infty} C_1 = E_{\mathbf{h}_{1,1}} \left[ \log\left( 1+ \frac{ || \mathbf{h}_{1,1} ||^2 P_1 }{ N_0 } \right) \right].
\end{equation}
As for the throughput of SU, although different channel realizations will activate different SUs, the overall throughput is still the ergodic capacity with respect to the distribution of $\mathbf{h}_{2,2}$ and $\mathbf{h}_{2,1}$, thanks to the independence of channels:
\begin{equation}
\label{eqn:CR_OSO_SIMO_N1_asympSU}
\lim_{K\rightarrow \infty} C_2 = E_{\mathbf{h}_{2,2},\mathbf{h}_{2,1}} \left[ \log\left( 1+ \frac{ || \mathbf{h}_{2,2} ||^2 P_2 }{ N_0 \left( 1 + \frac{| \langle \mathbf{h}_{2,2}, \mathbf{h}_{2,1}  \rangle|^2 P_1}{|| \mathbf{h}_{2,2} ||^2 N_0} \right) } \right) \right].
\end{equation}
The performance upper bound plotted in Fig.~\ref{fig:plot1} is the sum of (\ref{eqn:CR_OSO_SIMO_N1_asympPU}) and (\ref{eqn:CR_OSO_SIMO_N1_asympSU}).

\subsubsection{$N>1$}
\label{sec:CR_OSO_SIMO_N}

Intuitively, relaxing the constraint that at most one active SU is allowed would increase the system overall throughput while still maintaining PU's performance by setting a threshold on the total interference. With $N$ active SUs, the interference power becomes
\begin{equation}
\beta_{K}^{(N)} = \frac{ \sum_{n=2}^{N+1}|\langle \mathbf{h}_{1,n}^{\sf{min}}, \mathbf{h}_{1,1} \rangle|^2}{|| \mathbf{h}_{1,1} ||^2}
\end{equation}
where $\{ \mathbf{h}_{1,n}^{\sf{min}} \}_{n=2}^{N+1}$ generates the $N$ minimum $|\langle \mathbf{h}_{1,n}, \mathbf{h}_{1,1} \rangle|^2$ among $K$ candidate SUs, and the SINR of the primary link can be similarly computed as in (\ref{eqn:CR_OSO_SIMO_N1_MRC_SINR2}). For a given threshold   $\gamma_{\sf{thr}} = 10^{(\gamma_{\sf{thr}(dB)}/10)}$ and channel realizations, the maximal number of active users can be determined as
\begin{equation}
\label{eqn:CR_OSO_SIMO_N_ChooseN}
N^* = \arg \max_M{ \left\{ \min_{ \substack{ k_1, \cdots, k_M \in \left\{2, \cdots, K+1 \right\} \\ k_i \neq k_j, \forall i \neq j} } \frac{ \sum_{m=1}^{M}|\langle \mathbf{h}_{1,k_{m}}, \mathbf{h}_{1,1} \rangle|^2 }{|| \mathbf{h}_{1,1} ||^2} \leq \left(\gamma_{\sf{thr}} -1 \right) \frac{N_0}{P_2} \right\} }.
\end{equation}

\begin{figure}
\centering
\includegraphics[width=\textwidth]{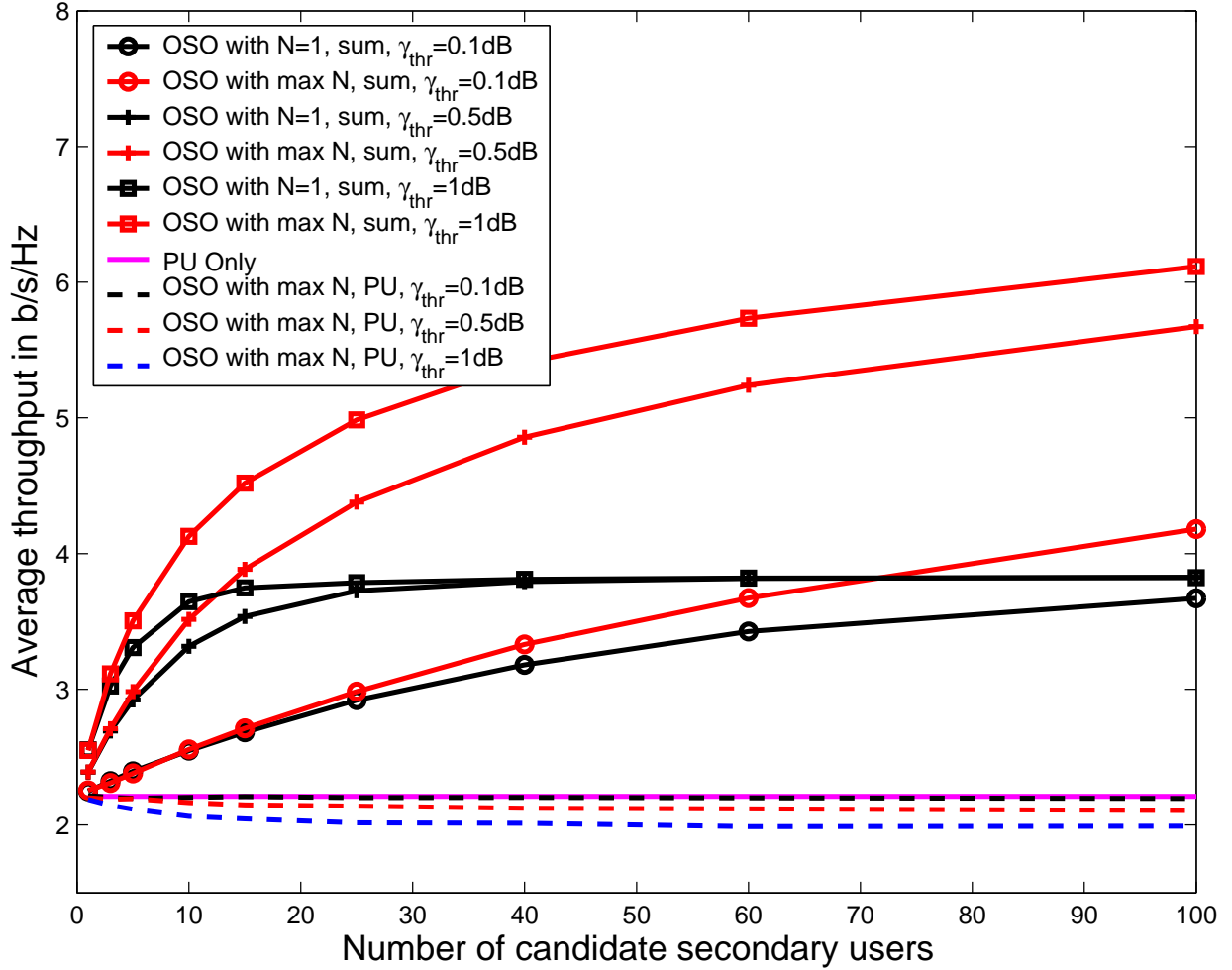}
\caption{Comparison of maximal $N$ and the special case $N=1$ in a SIMO system ($L_r=4$) with OSO. Throughput versus number of candidate SUs is plotted. Channels are i.i.d. Rayleigh fading. $\frac{P_1}{N_0} = \frac{P_2}{N_0} = 0$ dB. Three different interference threshold values are considered: $\gamma_{\textrm{$\sf{thr}$}} = 0.1, 0.5,$ and $1$ dB. PU's individual throughput in the OSO scheme is also plotted.}
\label{fig:plot3}
\end{figure}

Fig.~\ref{fig:plot3} compares the sum rate of maximal $N$ and the special case $N=1$. Apparently, allowing more secondary users whenever it is possible would further boost the sum rate, and this increase is much more remarkable when the interference margin is not stringent. Another interesting observation is the behavior of PU's throughput. Unlike the $N=1$ case in Fig.~\ref{fig:plot1}, where PU's throughput will converge to the no-interference case asymptotically, here due to the fact that we allow as many SUs as possible, asymptotically there is a constant gap between PU's throughput with and without interference. As the number of candidate SUs becomes large, there will be some SUs whose interference adds up to a level that approaches the interference margin $\gamma_{\textrm{$\sf{thr}$}}$. Hence this asymptotical gap is determined by the interference threshold.




\subsubsection{Remarks}
\label{sec:CR_OSO_SIMO_remarks}

\begin{remark}
Opportunistic Interference Alignment
\end{remark}

There is a very interesting interpretation of the OSO scheme with $N>1$. Recently  Interference Alignment (IA) \cite{CJ:08} has been proposed to deal with the problem that one receiver suffers interference from multiple sources. The key idea is that if the sources know their corresponding interference channel gains, they can perform precoding such that the resulting interference signals at the non-intended receiver only occupy a small number of dimensions. The essential idea of IA is quite similar to the proposed OSO scheme: force the interference signals to only point at a certain direction. The IA does so by allowing the non-intended transmitters to precode the signals, which requires a global knowledge of channel realizations. The proposed OSO scheme, however, achieves this goal by utilizing the randomness of the channels and relying on the multi-user diversity. The drawback of OSO is the requirement of multiple candidate SUs and increased workload of channel estimation, but the advantage is that there is no need for the SU transmitters to know the channel realizations, which is practically very difficult. From this perspective, the OSO scheme can be interpreted as \emph{Opportunistic Interference Alignment}.


\begin{remark}
Orthogonal Frequency-Division Multiple Access (OFDMA)
\end{remark}
Another advantage of the  OSO scheme is that \emph{it can be naturally incorporated into OFDMA systems}. This has great importance in practice as OFDMA is used extensively in modern wireless standards, such as the IEEE 802.16 mobile WiMAX, the 3GPP Long Term Evolution downlink, and the cognitive radio based IEEE 802.22. For a multi-user wireless network adopting OFDMA, the \emph{licensed carrier} owns all the frequency bands and allocates different subcarriers to different legitimate users in an OFDMA fashion. Consider the situation where the network is dense and hence all the subcarriers are occupied by legitimate users of the licensed carrier. Assume that there is an \emph{unlicensed carrier} who also wants to use these frequency bands to serve his subscribers with OFDMA. In this case, the OSO scheme can be directly applied in a subcarrier-by-subcarrier basis. On each subcarrier, there already exists one legitimate user, and the OSO scheme will find (if possible) one or more SUs of the unlicensed carrier whose interference is within the margin, provided that the number of candidate SUs is reasonably large. In this way, multiple SUs can be activated on different subcarriers and the overall throughput performance is improved without sacrificing the licensed carrier. Notice that the QoS of SUs cannot be guaranteed, and hence this scheme fits best for applications without stringent delay constraint and relatively low QoS requirement.

\begin{remark}
Comparison to previous multi-user diversity schemes
\end{remark}
Multi-user diversity was first recognized in \cite{KH:95,Tse:97,TH:98,VTL:02}, and has sparked intensive research interest in both academia and industry. The proposed OSO scheme also relies on this concept, but there is one important difference from previous works. In traditional multi-user systems, users are generally balanced. This is an important property for the development of multi-user diversity, as it requires some users to sacrifice their short-term throughput and wait for ``peaks'' to increase the long-term sum throughput. For any specific user, there is no performance guarantee  at any given time. This obviously cannot be directly applied to the CR system, since the PU is more important than SUs and his performance should be guaranteed. From this perspective, the proposed OSO scheme can be thought of as a multi-user diversity scheme for extremely asymmetric wireless networks.

\begin{remark}
The user fairness issue
\end{remark}
The proposed OSO scheme, just like other multi-user diversity schemes, has the user fairness issue if the static fading channel is considered. The primary receiver selects the secondary user(s) whose interference channel is orthogonal to the primary link. If the channel realization is very slowly varying, then one or several ``lucky'' SUs will be active for a long time, while other SUs remain silent. This creates the fairness problem among the candidate SUs.

The same problem was considered in the original multi-user diversity scheme \cite{VTL:02} and many following papers, and several solutions have been proposed in the  framework of \emph{user scheduling}. The general idea is to weigh users such that the ones who get to transmit  now have decreasing possibility of being selected again in the near future. One  example is the \emph{proportional fairness} principle in \cite{VTL:02}. However, one should notice that such user scheduling method will  not work in the OSO scheme. The reason is that user selection has to be based on the spatial orthogonalization. Hence, user scheduling can be only applied to the set of SUs who are orthogonal to the primary link, which is still unfair to the other candidate SUs.

There are some solutions attacking this problem. For example, as discussed in Remark 2, OFDMA in the wideband channel can help with the user fairness. Due to the independence of channel realizations in different subcarriers, the SU selection rule on each subcarrier is different. It is very unlikely that one user would satisfy most of them and correspondingly occupy a lot of subcarriers. What is more likely is that different SUs are activated on different subcarriers, which alleviates the user fairness issue.

Another solution is the \emph{random beamforming}, which relies on the artificially introduced randomness at the transmitter to vary the equivalent channel realization. Assume that each transmitter is also equipped with multiple antennas\footnote{The use of multiple transmit antennas here is to introduce randomness for user fairness. In Sec.~\ref{sec:CR_OSO_MIMO}, a different purpose of multiple transmit antennas will be elaborated.}. For the sake of simplicity, let us consider $L_t=L_r=2$. The scheme is to add a time-varying phase and power to the transmit antennas at PU-TX, similar to \cite[Figure 5]{VTL:02}. Note that without this random beamforming, the SIMO primary link channel is
\begin{equation}
\label{eqn:PU_chn_1}
\mathbf{h}_{1,1} = \left[ \begin{array}{l}
\mathbf{h}_{1,1}(1) \\
\mathbf{h}_{1,1}(2)
\end{array} \right]
\end{equation}
which remains constant over time and hence causes the user fairness issue. Now applying random multiplication factor $\sqrt{\alpha}$ on transmit antenna 1 and $\sqrt{1-\alpha} e^{j \theta}$ on transmit antenna 2, the equivalent SIMO channel of the primary link becomes
\begin{eqnarray}
\label{eqn:PU_chn_2}
\tilde{\mathbf{h}}_{1,1} &=& \mathbf{H}_{1,1} \left[ \begin{array}{c}
\sqrt{\alpha} \\
\sqrt{1-\alpha} e^{j \theta}
\end{array} \right] \nonumber \\
&=& \left[ \begin{array}{c}
\mathbf{H}_{1,1}(1,1) \sqrt{\alpha} + \mathbf{H}_{1,1}(1,2) \sqrt{1-\alpha} e^{j \theta}  \\
\mathbf{H}_{1,1}(2,1) \sqrt{\alpha} + \mathbf{H}_{1,1}(2,2) \sqrt{1-\alpha} e^{j \theta}
\end{array} \right].
\end{eqnarray}
By varying $\alpha$ and $\theta$ in a pseudo-random fashion, the equivalent SIMO channel also changes over time even when the physical channel is static, which will trigger different SUs to be activated at different times. Notice that such pseudo-randomness could be synchronized on both the transmitter and receiver side, which leads to two advantages. Firstly, the channel estimation needs to be done only once, with respect to the physical MIMO channel. Secondly, as long as the receiver gets the channel estimation of  $\mathbf{h}_{1,1}$ together with the interference links from candidate SUs, it can apply the pseudo-randomness to the primary link and perform SU selection with respect to each realization of $(\alpha, \theta)$, in an ``anti-causal'' way. That is to say, the SU selection does not have to wait until the favorable $(\alpha, \theta)$ appears. It can be done once the channel estimation is obtained, and the decision can be immediately fed back to the SU so it knows when to be active in the future.

\begin{remark}
Inferiority of the secondary users
\end{remark}
In the traditional CR  based on time orthogonalization, the SU's inferiority can be seen from the fact that it has to sense the absence of PU before transmitting, and once PU starts communication it has to stop transmitting. On the other hand, the transmission of SU will see an interference-free channel if PU is absent, which is advantageous.

For the OSO scheme, however, the SU's inferiority is exhibited in a different way. Now the selected SU(s) can transmit even if PU is present. From this perspective, SU is less inferior than the traditional CR. In contrast, the secondary link will typically see a strong-interference channel when transmitting, since PU's signal is not aligned at SU's receiver. This is where the SU is inferior to PU, whose receiver sees very little interference from SU.






\subsection{The OSO Scheme -- MIMO}
\label{sec:CR_OSO_MIMO}

The previous section only considers systems with single transmit and multiple receive antennas. The OSO scheme for this SIMO model is relatively simple, as the received signal at the intended receiver only spans one spatial dimension and hence opportunistic orthogonalization is easy to derive. From the capacity perspective, SIMO only provides receive SNR gain. It is well-known that using both multiple transmit and multiple receive antennas (full MIMO) will increase the capacity significantly thanks to the \emph{multiplexing gain}, which describes the additional spatial dimensions MIMO creates. This section extends the OSO scheme derive in Section~\ref{sec:CR_OSO_SIMO} to the MIMO channel, and incorporates SIMO as a special case. For simplicity of argument, we will focus on $L_t=L_r=2$ with $N=1$ and then discuss general $\left( L_t, L_r \right)$. Extension to general $N$ is straightforward.

\subsubsection{OSO based on MIMO eigen-beamforming}
\label{sec:CR_OSO_MIMO_BF}

The system model remains the same as in Section~\ref{sec:CR_model} and Fig.~\ref{fig:CR_model}, with the following two differences. The first is that now each transmitter also has $L_t>1$ antennas. The second is that we further assume each transmitter perfectly knows the channel between itself and PU1-RX, i.e., transmitter of the $n$-th user knows $\mathbf{H}_{1,n}$, $n=1, \cdots,N+1$. It should be noted that this is a valid assumption and many practical systems have either explicit\footnote{Note that PU1-RX is assumed to have perfect knowledge of $\left\{\mathbf{H}_{1,n} \right\}_{n=1}^{N+1}$ (see Section~\ref{sec:CR_model}). So it is possible for each transmitter $n$ to know $\mathbf{H}_{1,n}$ via CSI feedback.} or implicit feedback that allows transmitter to possess channel state information (CSIT). In practice this can be made easier if we limit the choice of $N$ to be small. For example, our main discussion is on $N=1$ and hence only two transmitters need to possess the corresponding CSIT: one PU and one active SU.

With these assumptions, the signal model of both receivers can be written as
\begin{eqnarray}
\textrm{Primary link}&:& \mathbf{y}_1 = \mathbf{H}_{1,1} \mathbf{F}_{1} \mathbf{x}_1 + \mathbf{H}_{1,2} \mathbf{F}_{2} \mathbf{x}_2 + \mathbf{z}_1; \label{eqn:CR_OSO_MIMO_N1_PU} \\
\textrm{Secondary link}&:& \mathbf{y}_2 = \mathbf{H}_{2,2} \mathbf{F}_{2} \mathbf{x}_2 + \mathbf{H}_{2,1} \mathbf{F}_{1} \mathbf{x}_1 + \mathbf{z}_2, \label{eqn:CR_OSO_MIMO_N1_SU}
\end{eqnarray}
where $\mathbf{H}_{mn}, \mathbf{F}_{n} \in \mathbb{C}^{2 \times 2}$, $\mathbf{x}_{m}, \mathbf{y}_{m}, \mathbf{z}_{m} \in \mathbb{C}^{2 \times 1}$, and $\mathbf{F}_{n}$ is the beamforming matrix at the $n$-th transmitter which is derived from channel matrix $\mathbf{H}_{1,n}$.

Since PU1-TX knows $\mathbf{H}_{1,1}$ perfectly, the capacity-optimal transmission policy without interference is to first perform the Singular Value Decomposition (SVD) of the channel matrix $\mathbf{H}_{1,1}$ \cite{Tel:95}
\begin{eqnarray}
\label{eqn:CR_OSO_MIMO_N1_SVD_H11}
\mathbf{H}_{1,1} &=& \mathbf{U}_{1,1} \mathbf{\Lambda}_{1,1} \mathbf{V}_{1,1}^{H} \nonumber \\
&=& \left[\mathbf{u}_{1,1}(1), \mathbf{u}_{1,1}(2) \right] \left[ \begin{array}{cc} \lambda_{1,1}(1) & 0 \\ 0 & \lambda_{1,1}(2) \end{array} \right] \left[\mathbf{v}_{1,1}(1), \mathbf{v}_{1,1}(2) \right]^{H}
\end{eqnarray}
to obtain the beamforming matrix
\begin{equation}
\mathbf{F}_{1} = \mathbf{V}_{1,1},
\end{equation}
where $\mathbf{U}_{1,1}, \mathbf{V}_{1,1} \in \mathbb{C}^{2 \times 2}$ are unitary matrices, and the diagonal matrix $\mathbf{\Lambda}_{1,1}$ consists of the ordered singular values $\left\{ \lambda_{1,1}(1) \geq \lambda_{1,1}(2) \right\}$. Together with left multiplying the received signal $\mathbf{U}_{1,1}^{H}$, the MIMO channel $\mathbf{H}_{1,1}$ is diagonalized, and then water-filling power allocation over the eigen-channels gives the optimal capacity \cite{Tel:95}.

The SVD-based beamforming together with water-filling power allocation over all eigen-channels is the theoretically optimal transmission scheme. In practice, however, beamforming using only a few strongest eigen-channels is commonly adopted. This scheme is suboptimal but has low complexity and can be easily implemented. In some scenarios it is in fact close to be optimal. For example, it is well-known that transmitting only on the strongest eigen-channel is the optimal strategy in the asymptotically low SNR regime. As another example, if the channel is \emph{ill-conditioned}, i.e., the condition number is very large, then only using a few strongest eigen-channels is also approximately optimal. This can be inferred from the water-filling power allocation. Recall that the principle of water-filling is to allocate more power if the eigen-channel is strong. Hence for the ill-conditioned channel, the matrix is close to be singular and the very weak eigen-channels will get very little power from water-filling. Consequently, discarding these eigen-channels incurs very little performance loss.

With this suboptimal scheme, both PU and SU will send only one data stream over one eigen-channel and the beamforming matrix $\mathbf{F}$ degenerates to a vector $\mathbf{f}$. The transmit beamforming of PU reduces to sending $x_1$ over the strongest eigen-channel $\left\{\lambda_{1,1}(1), \mathbf{v}_{1,1}(1)\right\}$:
\begin{equation}
\mathbf{f}_{1} = \mathbf{v}_{1,1}(1).
\end{equation}
Hence at PU1-RX,
\begin{eqnarray}
\mathbf{y}_1 &=& \mathbf{H}_{1,1} \mathbf{v}_{1,1}(1) x_1 + \mathbf{H}_{1,2} \mathbf{f}_{2} x_2 + \mathbf{z}_1 \nonumber \\
&=& \lambda_{1,1}(1) \mathbf{u}_{1,1}(1) x_1 + \mathbf{H}_{1,2} \mathbf{f}_{2} x_2 + \mathbf{z}_1.
\end{eqnarray}

Consider the SVD of $\mathbf{H}_{1,2}$:
\begin{eqnarray}
\label{eqn:CR_OSO_MIMO_N1_SVD_H12}
\mathbf{H}_{1,2} &=& \mathbf{U}_{1,2} \mathbf{\Lambda}_{1,2} \mathbf{V}_{1,2}^{H} \nonumber \\
&=& \left[\mathbf{u}_{1,2}(1), \mathbf{u}_{1,2}(2) \right] \left[ \begin{array}{cc} \lambda_{1,2}(1) & 0 \\ 0 & \lambda_{1,2}(2) \end{array} \right] \left[\mathbf{v}_{1,2}(1), \mathbf{v}_{1,2}(2) \right]^{H}.
\end{eqnarray}
Similar to the methodology in Section~\ref{sec:CR_OSO_SIMO}, we would like to exploit the channel randomness and independence to choose one SU whose \emph{weakest} eigen-channel $\mathbf{u}_{1,2}(2)$ is almost orthogonal to $\mathbf{u}_{1,1}(1)$. Two remarks are appropriate at this stage.
\begin{itemize}
\item [a)] We want to choose a beamforming direction of SU that is almost orthogonal to the space $\spa{\mathbf{u}_{1,1}(1)}$. This follows the same philosophy as in the SIMO case.
\item [b)] Item a) can be achieved if either of the two directions $\left\{\mathbf{u}_{1,2}(1), \mathbf{u}_{1,2}(2) \right\} $ is almost orthogonal to $\mathbf{u}_{1,1}(1)$. However, if the direction associated with the larger singular values $\lambda_{1,2}(1)$ is chosen, then any oscillation in $\mathbf{u}_{1,2}(1)$ will result in larger projection onto the space $\spa{\mathbf{u}_{1,1}(1)}$ and hence larger interference than choosing $\mathbf{u}_{1,2}(2)$ and $\lambda_{1,2}(2)$. This is especially true if the channel matrix $\mathbf{H}_{1,2}$ is ill-conditioned.
\end{itemize}

Since
\begin{equation}
\mathbf{f}_{2} = \mathbf{v}_{1,2}(2),
\end{equation}
the overall received signal at PU1-RX becomes
\begin{equation}
\label{eqn:CR_OSO_MIMO_N1_PU1-RX}
\mathbf{y}_1 = \lambda_{1,1}(1) \mathbf{u}_{1,1}(1) x_1 + \lambda_{1,2}(2) \mathbf{u}_{1,2}(2) x_2 + \mathbf{z}_1.
\end{equation}
This is the same signal model as in Equation~(\ref{eqn:CR_OSO_SIMO_N1_PU}). The OSO scheme discussed in Section~\ref{sec:CR_OSO_SIMO_N1} can be directly applied to this model. The SINR of PU can be written as
\begin{equation}
\label{eqn:CR_OSO_MIMO_N1_PU_SINR}
\textrm{$\sf{SINR}_1$} = \frac{ \lambda_{1,1}^2(1) P_1 }{ \lambda_{1,2}^2(2) | \langle \mathbf{u}_{1,1}(1), \mathbf{u}_{1,2}(2) \rangle |^2 P_2 + N_0}.
\end{equation}

There is a very interesting observation from Equation~(\ref{eqn:CR_OSO_MIMO_N1_PU_SINR}). As we have mentioned, one motivation for each user to use only one eigen-channel is the ill-conditioned channel matrix. In the $L_t=L_r=2$ MIMO channel, this means $ \lambda(1)$ is much larger than $ \lambda(2)$. This channel illness has two direct impacts on PU's SINR. First, the intended signal $x_1$ is sent on the strongest eigen-channel, and hence the signal power is amplified by $\lambda_{1,1}^2(1)$. Since $\mathbf{H}_{1,1}$ is ill-conditioned, $\lambda_{1,1}(1)$ is typically large, which is beneficial to $\textrm{$\sf{SINR}_1$}$. At the same time, the channel illness of $\mathbf{H}_{1,2}$ suggests that $\lambda_{1,2}(2)$ is very small, which is also beneficial to $\textrm{$\sf{SINR}_1$}$ as it further reduces the residual interference from the imperfect orthogonalization between $\mathbf{u}_{1,1}(1)$ and $\mathbf{u}_{1,1}(1)$. In other words, very small $\lambda_{1,2}(2)$ relaxes the requirement of orthogonalization for a given interference margin. In fact, consider the extreme case where $\mathbf{H}_{1,2}$ is \emph{singular}, which means that $\lambda_{1,2}(2) = 0$. In this extreme case, \emph{$\mathbf{u}_{1,1}(1)$ and $\mathbf{u}_{1,2}(2)$ do not need to be orthogonal at all}: the interference power $\lambda_{1,2}^2(2) | \langle \mathbf{u}_{1,1}(1), \mathbf{u}_{1,2}(2) \rangle |^2$ will be zero all the time no matter what value $\langle \mathbf{u}_{1,1}(1), \mathbf{u}_{1,2}(2) \rangle$ is. Such extreme example demonstrates the benefit of ill-conditioned MIMO channels in the OSO scheme.

This extreme case motivates a simple two-stage \emph{secondary user selection algorithm}: Check the rank of all $\left\{\mathbf{H}_{1,n} \right\}_{n=2}^{K}$ in the first stage. If there is one (or more) $\mathbf{H}_{1,n}$ which is singular, the corresponding SU(s) are activated. If such SU(s) cannot be found, move on to the second stage and use the same procedure as in Section~\ref{sec:CR_OSO_SIMO_N1} to evaluate the interference power of each candidate SU and compare it with the threshold $\gamma_{\textrm{$\sf{thr}$}}$ to determine whether it can be activated or not.

On the other hand, the overall received signal at SU2-RX is
\begin{equation}
\label{eqn:CR_OSO_MIMO_N1_SU_RX}
\mathbf{y}_2 = \mathbf{H}_{2,2} \mathbf{v}_{1,2}(2) x_2 + \mathbf{H}_{2,1} \mathbf{v}_{1,1}(1) x_1 + \mathbf{z}_2.
\end{equation}
A second interesting observation can be made from Equation~(\ref{eqn:CR_OSO_MIMO_N1_SU_RX}) and the independence between $\left\{\mathbf{H}_{m,n}, m,n=1,2 \right\}$. Notice that $\mathbf{v}_{1,2}(2)$ indicates the weakest direction of $\mathbf{H}_{1,2}$, and due to the independence between $\mathbf{H}_{1,2}$ and $\mathbf{H}_{2,2}$ it is highly impossible for $\mathbf{v}_{1,2}(2)$ to be aligned to either the strong or weak eigen-channel of $\mathbf{H}_{2,2}$. Hence, although the transmitter signal $x_2$ is aligned to be orthogonal to the intended signal at PU1-RX, no signal alignment takes place at SU2-RX. Meanwhile, $x_1$ from PU becomes the interference signal at SU2-RX, and since $\mathbf{v}_{1,1}(1)$ is  the strongest eigen-channel of $\mathbf{H}_{1,1}$, it will be independent of $\mathbf{H}_{2,1}$ and hence there is no interference control for SU. This remains the same as in the SIMO case.

\begin{figure}
\centering
\includegraphics[width=\textwidth]{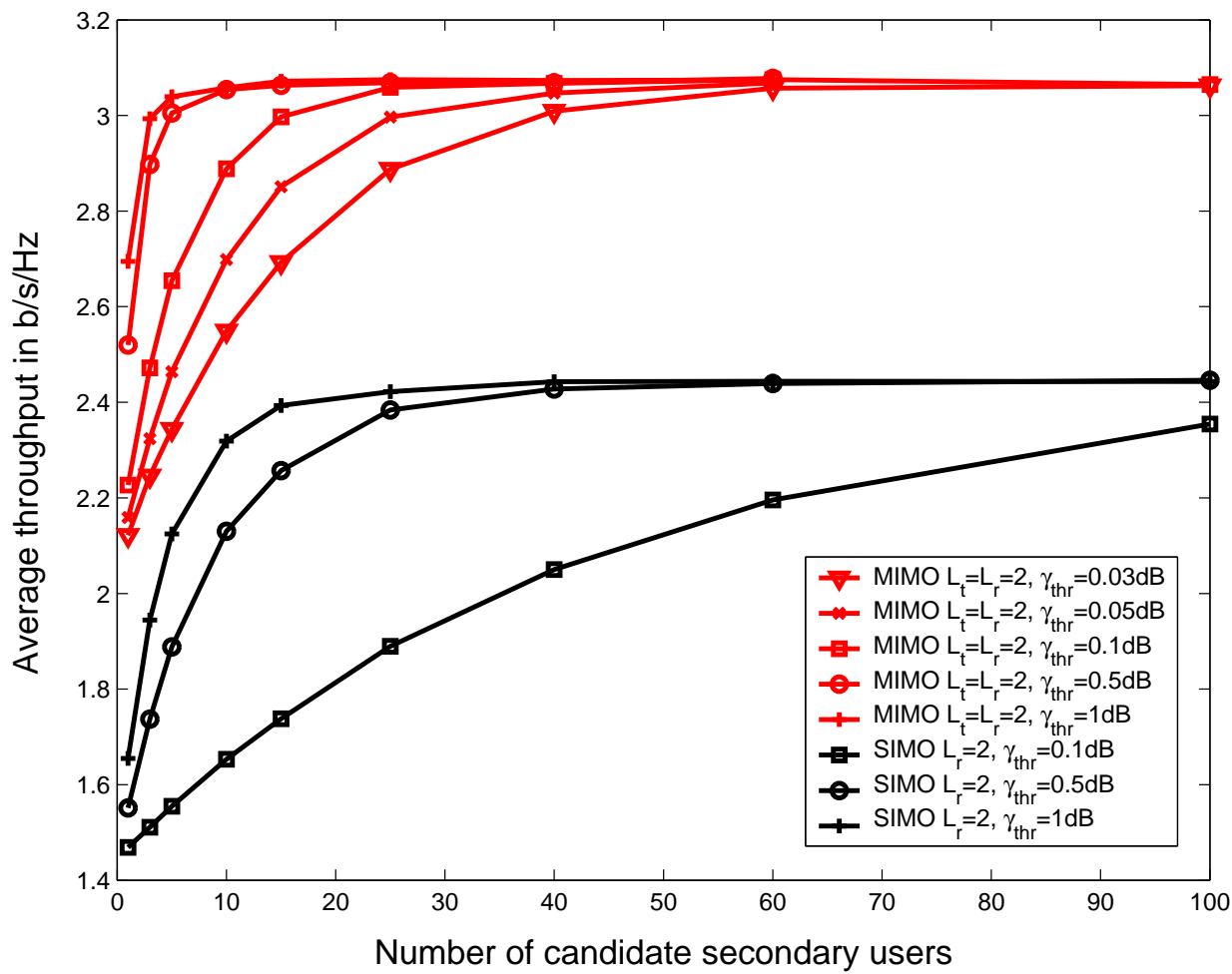}
\caption{Comparison of the average sum capacity for the OSO scheme in both MIMO ($L_t=L_r=2$) and SIMO ($L_t=1,L_r=2$) cognitive radio systems. Sum throughput versus number of candidate SUs is plotted. Maximum number of active SUs is $N=1$. Channels are i.i.d. Rayleigh fading. $\frac{P_1}{N_0} = \frac{P_2}{N_0} = 0$ dB. Interference threshold values are indicated in the figure.}
\label{fig:plot4}
\end{figure}

Fig.~\ref{fig:plot4} gives the sum throughput performance of the derived OSO scheme in a $L_t=L_r=2$ MIMO CR. For comparison the sum throughput of SIMO $L_t=1,L_r=2$ is also plotted. The important observation is that even with the same signal and noise power, the sum throughput of MIMO is significantly larger than SIMO. \emph{This should not be explained as MIMO provides more multiplexing gain than SIMO}: even MIMO uses only one spatial dimensions for data transmission of each user. The throughput advantage is best understood by comparing the MIMO Equation~(\ref{eqn:CR_OSO_MIMO_N1_PU_SINR}) with the SIMO Equation~(\ref{eqn:CR_OSO_SIMO_N1_MRC_SINR1}). Let us rewrite Equation~(\ref{eqn:CR_OSO_SIMO_N1_MRC_SINR1}) as
\begin{equation}
\textrm{$\sf{SINR}_1$} = \frac{ || \mathbf{h}_{1,1} ||^2 P_1 }{ || \mathbf{h}_{1,2} ||^2 |\langle \frac{\mathbf{h}_{1,2}}{||\mathbf{h}_{1,2}||}, \frac{\mathbf{h}_{1,1}}{||\mathbf{h}_{1,1}||} \rangle|^2 P_2 +N_0}. \label{eqn:CR_OSO_SIMO_N1_MRC_SINR11}
\end{equation}
Compared to (\ref{eqn:CR_OSO_MIMO_N1_PU_SINR}), we can see that $|| \mathbf{h}_{1,1} ||^2$ ($|| \mathbf{h}_{1,2} ||^2$) plays the role of $\lambda_{1,1}^2(1)$ ($\lambda_{1,2}^2(2)$). The typical value of $|| \mathbf{h}_{m,n} ||^2$ is the average channel power, and  hence roughly speaking, $|| \mathbf{h}_{1,1} ||^2 \simeq || \mathbf{h}_{1,2} ||^2$ since $\mathbf{h}_{1,1}$ and $\mathbf{h}_{1,2}$ are i.i.d. On the other hand, $\lambda_{1,1}(1)$/$\lambda_{1,2}(2)$ are the largest/smallest singular values of $\mathbf{H}_{1,1}$/$\mathbf{H}_{1,2}$, respectively, and  typically $\lambda_{1,1}^2(1)$ is much larger than $\lambda_{1,2}^2(2)$. This is why the OSO scheme for MIMO leads to much better average throughput than the SIMO case.


\subsubsection{Multi-user diversity on the MIMO eigen-channels}
\label{sec:CR_OSO_MIMO_MUD}


%

The previous observation can be generalized as the multi-user diversity on the MIMO eigen-channels, which is explained in the following. Let us go back to the $L_t=L_r=2$ MIMO channels $\mathbf{H}_{1,1}$ and $\mathbf{H}_{1,2}$. The philosophy of OSO is to choose the SU whose $\mathbf{u}_{1,2}(2)$ is almost orthogonal to $\mathbf{u}_{1,1}(1)$. For the sake of argument we assume the  perfect alignment where $\langle \mathbf{u}_{1,2}(2) ,\mathbf{u}_{1,1}(1)\rangle = 0$.

\begin{figure}
\centering
\includegraphics[width=\textwidth]{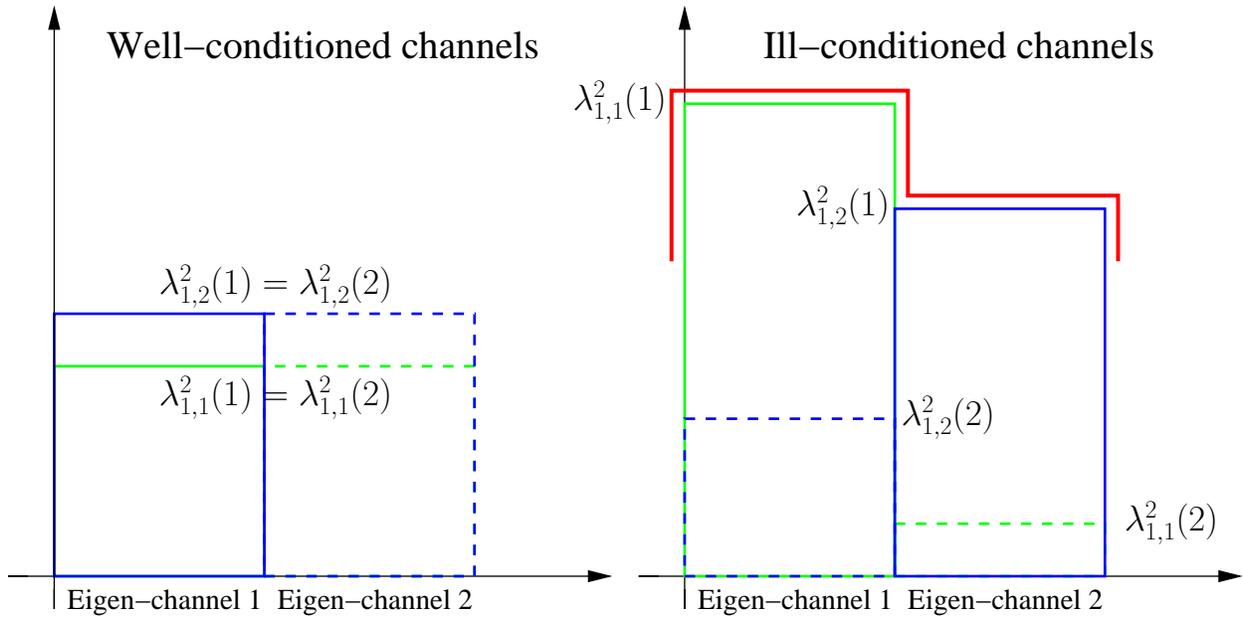}
\caption{Illustration of the multi-user diversity on the $L_t=L_r=2$ MIMO eigen-channels with two users. User 1 is plotted with the green curves and user 2 is with the blue ones. Solid and dashed curves correspond to strong and weak eigen-channels, respectively. The plot on the left is for the well-conditioned channels, in which the total channel power is  equally distributed over eigen-channels. The plot on the right shows the multi-user diversity effect for the ill-conditioned channels. The red curve illustrates the effect of ``riding the peaks''.}
\label{fig:plot5}
\end{figure}

From a point-to-point perspective, well-conditioned MIMO channel matrix is beneficial as it allows more degrees of freedom for communication and improves the performance of linear detectors such as Zero-Forcing (ZF) and Minimum Mean Square Error (MMSE) \cite[Chapter 7]{TV:05}. This view is shifted in the multi-user case, which is illustrated in Fig.~\ref{fig:plot5}. If both channels are well-conditioned, there are no obvious ``peaks'' to exploit and the multi-user diversity gain is marginal, as is demonstrated in  the left plot of Fig.~\ref{fig:plot5}. On the other hand, ill-conditioned channel matrix can greatly improve the overall throughput by ``riding the peaks'', which is depicted in the right plot of Fig.~\ref{fig:plot5}. Recall that SU's weak channel is aligned to PU's strong channel. This allows both users to beam on their strong channels without interfering with each other. This sum throughput advantage is depicted with the red curves in Fig.~\ref{fig:plot5}.


Note that whether the MIMO channel is well- or ill-conditioned not only decides the multi-user diversity gain, but determines the admission of SUs into the cognitive radio system. At the same time, channel conditions (e.g., rank) are determined from physics and cannot be controlled by system design. As has been emphasized, the CR system is unbalanced and PU's performance should be strictly guaranteed. Hence one can formulate a natural adaptation scheme for CR system based on OSO as follows. If PU's channel is well-conditioned, it should use all the spatial dimensions for data transmission, which leaves no room for SU to exploit additional spatial dimensions. In this scenario, there is very little incentive for PU to use only one spatial dimension and leave the other to SU, because this will affect PU's performance significantly. On the other hand, if PU's channel is extremely ill-conditioned, there is very small loss if PU only uses the strong eigen-channel. In this case, the overall throughput will be increased significantly, thanks to the multi-user diversity gain, but PU's individual performance remains almost undamaged.




\subsubsection{Beyond MIMO eigen-beamforming}
\label{sec:CR_OSO_MIMO_MoreLtLr}
The OSO scheme discussed in Section~\ref{sec:CR_OSO_MIMO_BF} assumes $L_t=L_r=2$ and $N=1$ for simplicity of discussion. If there are more antennas, the spatial dimensions will be increased and there is more room for the SUs to co-exist with PU. For example, consider a $L_t=L_r=4$ MIMO system. The PU might only use the two strongest eigen-channels and is already approaching the optimal performance. This leaves two other spatial dimensions for the SUs, who can take advantage by relying on multi-user diversity and opportunistic interference alignment. In summary, more antennas creates more spatial dimensions, and there is a better chance that some of them are so weak that PU's performance will not degrade much if they are unused. This gives a better opportunity for SUs to co-exist and improves the overall system throughput.


Similar analysis based on interference threshold can be performed, although the procedure will become more complicated. The complication comes from the fact that PU has multiple independent streams, and SU's interference on each of them needs to be considered. A global threshold on the interference power is hence difficult to obtain. Instead we should directly study the decrease of MIMO capacity due to the cross interference. For example, an optimization problem can be formulated where the objective function is the (weighted) sum rate of both PU and SU, and the constraints are PU's QoS in addition to transmit power. However, one should be very careful with such approach, as the resulting optimal beamforming matrices will depend on the global CSI $\left\{ \mathbf{H}_{i,j} \right\}$.

%



\section{Conclusion}
\label{sec:conc}

Opportunistic Spatial Orthogonalization (OSO) is a novel CR scheme that allows the existence of secondary users even when the primary user occupies all the frequency bands all the time. This scheme relies on the randomness and independence among channel matrices and exploits the spatial dimensions to orthogonalize users and hence minimize interference. The OSO scheme developed for SIMO has led to the interesting concept of opportunistic interference alignment. For the MIMO case, it is shown that ill-conditioned MIMO channel significantly increases the total throughput without much sacrifice of PU's performance.

The general concept behind OSO can be applied to a broad class of multi-user systems other than CR, and for some of them, we believe the throughput gain will be even more prominent. CR is an unbalanced system, and PU's performance needs to be strictly guaranteed. This constraint in fact prevents us from fully exploiting the benefit of OSO. For example, we have discussed in Section~\ref{sec:CR_OSO_MIMO_MUD} that if the channel is well-conditioned, there is almost no incentive for PU to give up one dimension for SU to exploit. This certainly limits our exploitation in multi-user diversity. Without this constraint, however, even if both users have well-conditioned MIMO channels and hence very little variation across eigen-channels, we can \emph{artificially} add such variation by allocating different power on the eigen-channels, which will create larger ``peaks'' for multi-user diversity. This is similar to the opportunistic beamforming idea in \cite{VTL:02}. It is obvious that such scheme cannot guarantee the performance of any individual user, and hence is prohibited in the CR setting. However, from the sum throughput perspective, it is highly advantageous.


\section*{Acknowledgment}
Cong Shen would like to thank Wenyi Zhang and Ahmed Sadek from Qualcomm, and Tie Liu from Texas A\&M University for helpful discussions.

\bibliographystyle{IEEEtran}
\bibliography{Opportunistic}

\newpage

\end{document}